\documentstyle[12pt,epsfig]{article}
\begin{document}
\thispagestyle{empty}

\hfill{DOE/ER/40561-342-INT97-00-181}

\hfill{DOE/ER/41014-37-N97}
\vfill
\vspace*{24pt}
\begin{center}
{\large\bf Nucleon-Deuteron Scattering from an Effective Field Theory}

\vspace*{48pt}

{\bf P.F. Bedaque$^a$} and {\bf U. van Kolck$^b$\footnote{
Address after Jan 1 1998: W.K. Kellogg Radiation Laboratory, 
Caltech, Pasadena, CA 91125}}

\vspace*{12pt}

{\sl $^a$Institute for Nuclear Theory and $^b$Department of Physics}\\
{\sl University of Washington}\\
{\sl Seattle, WA 98195-1560}

e-mails: bedaque, vankolck@phys.washington.edu

\vspace*{36pt}

\begin{abstract}
We use an effective field theory  to 
compute low-energy nucleon-deuteron scattering.
We obtain the quartet scattering length using low energy constants 
entirely determined from low-energy nucleon-nucleon scattering.
We find $a_{th}=6.33$ fm, to be compared to $a_{exp}=6.35\pm 0.02$ fm.

\end{abstract}

\vspace*{12pt}

\end{center}

\vfill
\newpage
\setcounter{page}{1}

There has been considerable interest lately in a description of 
nuclear forces from the low-energy effective field theory (EFT) of QCD.
(For a review, see Ref. \cite{bira0}.)
Following a program suggested by Weinberg \cite{weinberg}, 
the leading components of the nuclear potential have been derived \cite{bira1} 
and a reasonable fit to two-nucleon properties has been achieved \cite{bira2}.
The correct formulation of the nuclear force problem within the EFT method is
important because it will allow a systematic calculation of nuclear 
properties consistently with QCD. One would like, for example, to devise
a theory of nuclear matter rooted in a hadronic theory that treats chiral 
symmetry correctly and yields the well-known few-nucleon phenomenology. 
One hopes that after a number of parameters of the EFT are either 
calculated from first principles or fitted to a set of few-nucleon data, 
the theory
can be used to predict other reactions involving light nuclei
and features of heavier nuclei.

However, some issues concerning renormalization in this non-perturbative
context and fine-tuning in the two-nucleon $S$-waves have been raised in
Refs. \cite{david1,david2} and are still not fully understood \cite{lots}.
The fine-tuning
necessary to bring a (real or virtual) bound state very close to
threshold generates a scattering length $a$ much larger than
other scales in the problem. At momenta of $O(1/a)$,
mesons can be integrated out and
the characteristic mass scale $\mu$ of the underlying theory
controls the size of the other effective range parameters;
for example, 
the effective range $r_0\sim 2/\mu$.
Once the leading order contributions, which give rise to $a$,
are included to all orders,
the EFT at momenta $O(1/a)$
becomes an expansion in powers of $1/(a \mu)$.
Kaplan \cite{david2} has noticed that the 
interactions that generate a non-zero $r_0$ can also be resummed
by the introduction of a baryon number two 
state 
of mass $\Delta= 2/M a |r_0|$, which in
lowest order in a derivative expansion couples to two nucleons with
a strength $g^2/4\pi= 1/M^2 |r_0|$.
In Ref. \cite{david2} it was shown how this works
in the two-nucleon $^1S_0$ channel.
Analogous considerations hold for the $^3S_1$ channel,
where they are similar to the old quasi-particle 
approach of Weinberg \cite{weinbergqp}.

In this paper we consider the application of these ideas to the 
three-nucleon system. Our goal here is to calculate some of the 
three-nucleon parameters that are dominated by the leading interactions
in the EFT without pions. 
We show in particular that 
the quartet scattering length in neutron-deuteron scattering can be 
predicted once the EFT is constrained by low-energy two-nucleon data.
Such an attempt to a model-independent or ``universal'' approach is 
not a new idea; it permeates for example the work of
Efimov (see, e.g., \cite{efimov}) and Amado (see, e.g., \cite{amante}). 
However, as we will show,
the EFT formulation is much easier to implement, from both 
conceptual and practical standpoints.  

\newcommand{\boldtau}{\mbox{\boldmath $\tau$}}
\newcommand{\boldT}{\mbox{\boldmath $T$}}

For momenta of order $1/a$ (the momentum scale relevant for zero-energy 
$Nd$ scattering),
we can integrate out mesons 
and consider an EFT with 
only nucleons $N$. Interactions are then described by
a tower of nucleon contact operators with an increasing number of 
derivatives. 
Amplitudes in leading order are given by a zero-range four-nucleon
interaction iterated 
to all orders. Corrections come in powers of $1/(a m_\pi)$.
The next two orders in this expansion, $1/(a m_\pi)\sim r_0/2a$ and 
$1/(a m_\pi)^2\sim (r_0/2a)^2$, 
stem from one and two insertions of 
a two-derivative four-nucleon operator giving rise to a non-zero $r_0$.
It is advantageous to sum
all the contributions coming from this operator, 
which can be easily done because
they appear in a geometric series. 
The resulting interaction is
equivalent to the s-channel propagation of a dibaryon, and therefore
can be obtained more directly by the introduction of a dibaryon field.
Assuming naturalness, only
higher orders depend on further $NN$ scattering parameters 
---such as the shape parameter, which contributes at 
$O(1/(a m_\pi)^3)$--- and
three-nucleon forces, which start at $O(1/(a m_\pi)^4)$.

Since in both $I=0$ and $I=1$ $S$-wave two-nucleon channels we observe
(one real, one virtual) bound states near threshold, we consider 
two dibaryon fields, 
$\boldT$ ($\vec{D}$) of spin zero (one) and isospin one (zero). 
The most general Lagrangian invariant
under parity, time-reversal, and small Lorentz boosts is
\begin{eqnarray}
\cal L & = & N^\dagger(i\partial_{0}+\frac{\vec{\nabla}^{2}}{2M}+\ldots)N 
                                                               \nonumber \\
  &  & + \boldT^\dagger\cdot(-i\partial_{0}-\frac{\vec{\nabla}^{2}}{4M}
                             +\Delta_T+\ldots)\boldT
            + \vec{D}^\dagger\cdot(-i\partial_{0}-\frac{\vec{\nabla}^{2}}{4M}
                              +\Delta_D+\ldots)\vec{D}           \nonumber \\
  &  & -\frac{g_T}{2} (\boldT^\dagger\cdot N\sigma_2\boldtau\tau_2N 
                       +\mbox{h.c.})
       -\frac{g_D}{2} (\vec{D}^\dagger\cdot N\tau_2\vec{\sigma}\sigma_2N 
                       +\mbox{h.c.})
       +\ldots                       \label{lag}
\end{eqnarray}
\noindent
Here the $\Delta_{T,D}$ and $g_{T,D}$ are undetermined parameters
and ``$\ldots$'' stands for higher order terms.
(Note that the effects of non-derivative and two-derivative four-nucleon terms 
can be absorbed into a redefinition of $\Delta_{T,D}$ and $g_{T,D}$ and
higher order four-nucleon terms.) 

\begin{figure}[t]
\centerline{\epsfig{file=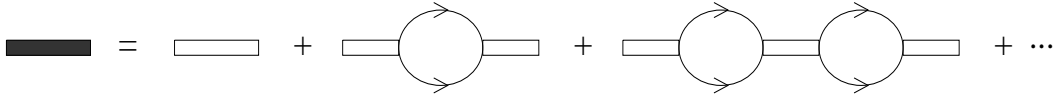,height=.5in,width=5.5in}}
\caption{Dressed dibaryon propagator.}
\label{fig1}
\end{figure}

In this non-relativistic theory all particles propagate forward in time,
nucleon tadpoles vanish and,
as a consequence, there is no dressing of the
nucleon propagator, which  is simply
\begin{equation}
S_N(p) = {i\over  p^0- \frac{\vec{p}^{\,2}}{2M} +i\epsilon}. \label{Nprop}
\end{equation}
\noindent
The propagators for dibaryons are more complicated, because of the
coupling to two-nucleon states. 
The dressed propagators consist of the bubble sum 
in Fig. \ref{fig1}, which amounts to a self-energy contribution 
proportional to the bubble integral. This integral is 
proportional to the (large) mass $M$, and it is this enhancement that 
gives rise to non-perturbative phenomena and leads eventually to the 
existence of bound states. 
The integral is also ultraviolet 
divergent and requires regularization. Introducing a cut-off $\Lambda$
we find a linear divergence $\propto \Lambda$, 
a cut-off independent piece which is non-analytic in the energy,
a term that goes as $\Lambda^{-1}$ and terms that are higher order in 
$\Lambda^{-1}$. The first and third terms can be absorbed in 
renormalization of the parameters of the Lagrangian (\ref{lag}); 
in what follows we omit a label $R$ that should be attached to these
parameters, i.e.,
$\Delta_{T,D}$ and $g_{T,D}$  stand for the renormalized parameters.
Higher order terms are neglected because they are of the same
order as interactions in the ``$\ldots$'' of the Lagrangian (\ref{lag}).
A dibaryon propagator has therefore the form
\begin{equation}
i S_D(p) =  \frac {1}{p^0- \frac{\vec{p}^{\,2}}{4M} - \Delta_D
             + \frac{M^2 g_D^{2}}{2\pi} 
               \sqrt{-M p^0+\frac{\vec{p}^{\,2}}{4}-i\epsilon} +i\epsilon} .
                                   \label{Dprop}
\end{equation}
\noindent
Note that such a dressed propagator has two poles at
$p^0= \vec{p}^{\,2}/4M-B$, $p^0= \vec{p}^{\,2}/4M -B_{deep}$ 
and a cut along the positive real axis
starting at $p^0= \vec{p}^{\,2}/4M$. 

\begin{figure}[t]
\centerline{\epsfig{file=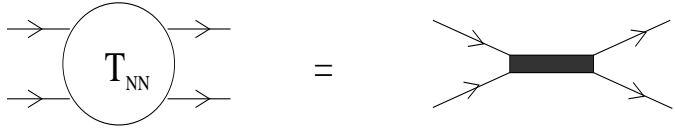,height=.65in,width=3.5in}}
\caption{$NN$ amplitude.}
\label{fig2}
\end{figure}

The $NN$ amplitude can now be obtained directly from $S_D(p)$
as in Fig. \ref{fig2}. 
In the center-of-mass, the on-shell $I=0$, $J=1$ $S$-wave amplitude 
at an energy $E= k^2/M$ is 
\begin{equation}
^{3}T_{NN}(k) = {4 \pi \over M} 
                    {1\over -\frac{2 \pi \Delta_D}
                                  {M g_D^{2}} 
                    +\frac{2 \pi }
                          {M^2 g_D^{2}}k^2
                    -i k},      \label{NNamp}
\end{equation}
\noindent
which is exactly equivalent to the effective range expansion.
An analogous result holds for the $I=1$ $S$-wave.
The four parameters $\Delta_{T,D}$ and $g_{T,D}$ can then be fixed from 
the experimentally known scattering lengths and effective ranges.
The $NN$ amplitude has shallow poles at 
$B\sim 1/M a^2$ which are associated with the deuteron 
in the $^3S_1$ channel and with the virtual bound state in the $^1S_0$
channel. The effective theory has also an additional deep bound state 
in each channel at
$B_{deep}\sim 4/M r_0^2$, which is outside the range of validity of the EFT.

{}From the triplet parameters $^3a= 5.42$ fm and $^3r_0=1.75$ fm
\cite{nijm} we find $\Delta_D =8.7$ MeV and $g_D^2 = 1.6 \cdot 10^{-3}$ MeV.
The resulting deuteron binding energy is $B= 2.28$ MeV.
{}From the singlet parameters 
$^1a_{pp}=-17.3$ fm, $^1a_{np}=-23.75$ fm, $^1a_{nn}=-18.8$ fm,
$^1r_{0 \, pp}=2.85$ fm, $^1r_{0 \, np}=2.75$ fm, and $^1r_{0 \, nn}=2.75$ fm
\cite{jerry} we find the averages 
$\Delta_T = -1.5$ MeV and $g_T^2 = 1.0 \cdot 10^{-3}$ MeV.

\begin{figure}[t]
\centerline{\epsfig{file=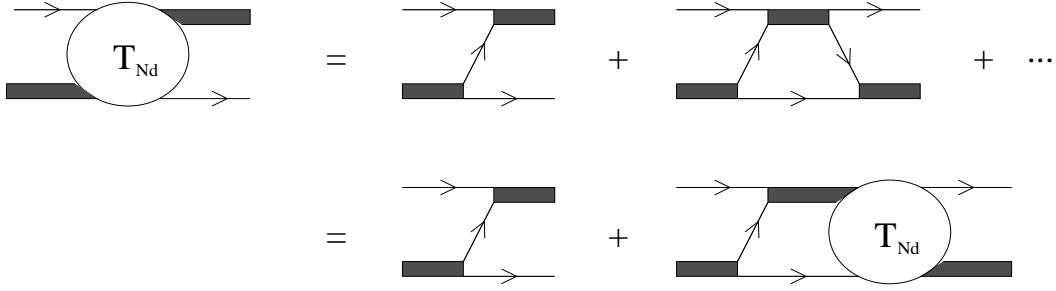,height=1.5in,width=5.5in}}
\caption{$Nd$ amplitude.}
\label{fig3}
\end{figure}

With the parameters so determined, we turn now to possible 
predictions in low-energy nucleon-deuteron scattering.
For simplicity we restrict ourselves to 
scattering below the deuteron break-up threshold, 
where the $S$-wave is dominant.
There are two $S$-wave channels, corresponding
to total spin $J=3/2$ and $J=1/2$. 
In the quartet only $\vec{D}$ contributes 
while in the doublet $\boldT$ also appears.
The $Nd$ scattering amplitude $T_{Nd}$ from the same interactions
is given by the diagrams in Fig. \ref{fig3}, 
which can be summed up by solving an integral equation in the quartet
and a pair of coupled integral equations in the doublet channel.
The diagrams under consideration are power-counting finite, but this does 
not preclude the existence of relevant contact interactions between
the nucleon and the dibaryons. 
Pion exchange that would generate such interactions can be expected to be 
larger for the $I=1$ dibaryon $\boldT$, and therefore predominantly
affect the $J=1/2$ channel. We will return to the doublet case in a future 
publication. Here we study the quartet 
channel expected to be much less sensitive
to the details of the physics of momenta of $O(m_\pi)$,
since the wave function on this channel vanishes by symmetry  
when the three particles are at the same point.

An enormous simplification comes about  because the s-channel interaction
due to the dibaryon is both local and separable. This allows us to
write a simple integral equation that sums all the graphs in Fig. \ref{fig3}.
Performing the integration over the time-component of the loop 4-momentum,
we find that the conveniently normalized
on-shell amplitude as a function of the initial (final) 
center-of-mass 3-momentum $\vec{k}$ ($\vec{p}$) satisfies
\begin{eqnarray}
\lefteqn{ \left[\frac{3(\vec{p}^{\, 2}-\vec{k}^2) }
                {8 M^2 g_D^2}
           +\frac{1}
                 {4 \pi}({\sqrt{\frac{3}
                                          {4}(\vec{p}^{\, 2}-\vec{k}^2)+M B}
                             -\sqrt{M B}})\right]
 \frac{t(\vec{p}, \vec{k})}
      {\vec{p}^{\, 2}-\vec{k}^2-i \epsilon }}    \label{aeq} \\
& & = \frac{-1}
           {(\vec{p}+\vec{k}/2)^2+M B}
       -   \int \frac{d^3 l}
                             {(2\pi)^3} 
       \frac{1}
            {\vec{l}^2 +\vec{l}\cdot\vec{p}+\vec{p}^{\, 2}-\frac{3}
                                                           {4}\vec{k}^2+M B}
         \frac{t(\vec{l}, \vec{k})}
              {\vec{l}^2-\vec{k}^2-i\epsilon}.    \nonumber 
\end{eqnarray}
\noindent
Note that all terms in a perturbative expansion of $t$ in (\ref{aeq})
are of the same order ($\sim 1/\sqrt{M B}$).
It is straightforward but tedious to show that the wave function
$(2\pi)^3 \delta(\vec{p}-\vec{k}) 
+  t(\vec{p}, \vec{k})/(\vec{p}^{\, 2}-\vec{k}^2 -i\epsilon) $
corresponding to a scattering solution indeed
satisfies the Schr\"odinger equation derived from the Lagrangian 
(\ref{lag}).

At zero energy ($k\rightarrow 0$) only the $S$-wave, depending on 
the magnitudes of momenta, contributes to the scattering, 
and we can perform the angular integration directly. It is also convenient 
to normalize all quantities to $\sqrt{M B}$. 
Defining 
$\vec{x}=\vec{p}/\sqrt{M B}$, 
\begin{equation} 
a(x)= \frac{\sqrt{M B}}{4 \pi} t(\frac{p}{\sqrt{M B}},0),         \label{adef}
\end{equation}
\noindent
and introducing 
\begin{equation}
F(x, z)= \frac{1}{xz} {\rm ln}(\frac{x^2+z^2+1+xz}
                                        {x^2+z^2+1-xz}),
\end{equation}
\noindent
Eq. (\ref{aeq})  becomes
\begin{equation}
{\frac{3}{4} \left[-\eta
 +\frac{1}{1+\sqrt{1+\frac{3}{4}x^2}}\right]
 a(x) } = -\frac{1}{x^2+1}
 -\frac{1}{\pi}\int_0^\infty dzF(x,z) a(z). 
\label{saeq}
\end{equation}
\noindent
Note that there is only one parameter
$\eta= 2\pi\sqrt{M B}/M^2 g_D^2= ^3\!\!r_0 \sqrt{M B}/2
       = 0.40$ in this equation.
The value of the function $a(x)$ at $x=0$ gives the $Nd$ scattering length in
units of $1/\sqrt{MB}$.
The same equation was previously  obtained and solved in the 
zero-range limit ($\eta\rightarrow 0$) \cite{skorny}.

\begin{figure}[t]
\vspace{-4.0cm}
\centerline{\epsfig{file=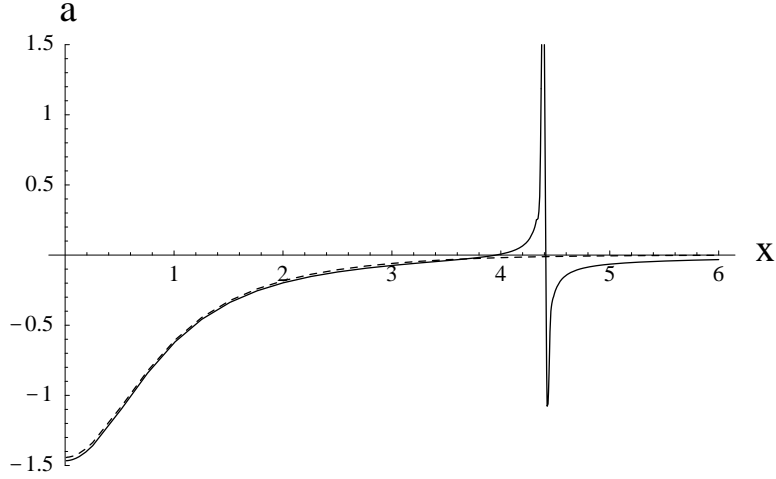,height=5.5in,width=4.0in}}
\vspace{-3.5cm}
\caption{Function $a(x)$ for $\eta =0.40$ without (solid line) and 
with (dashed line) cut-off.}
\label{fig4}
\end{figure}

We have solved Eq. (\ref{saeq}) numerically for $\eta= 0.40$ by the
Nystrom method \cite{recipes}.  
The solution $a(x)$ is plotted as the solid line in Fig. \ref{fig4}.
The pole in $a(x)$ around $x\sim 4.4$ is associated with the spurious deep 
two-body pole. Its presence allows intermediate states
where two nucleons 
fall into this deep state while 
the other has extra energy. 
This means that the outgoing 
wave has an additional component,
a pole at the momentum corresponding to this additional process. 
The interesting point is that even though 
the effective theory makes nonsensical predictions outside its 
domain of validity, like the existence of this new state in $Nd$ scattering, 
the low-$x$ part of the curve is insensitive to the large-$x$
behavior, and the prediction for the scattering length is sensible.
In order to demonstrate this more explicitly we have also solved
Eq. (\ref{saeq}) with a cut-off two-nucleon amplitude without the deep
pole. For a cut-off of 150 MeV we obtain the 
broken line in Fig. \ref{fig4}. 

The quartet scattering length is $^4a= - a(0)/\sqrt{M B}$.
For $\eta=0$ (and $B$ fixed), we reproduce the result 
$^4a=5.09$ fm of Ref. \cite{skorny}.
Taking into account the finite range ($\eta=0.40$)  we  obtain 
(Fig. \ref{fig4}) $^4a=6.33$ fm with an uncertainty from higher orders of 
$\sim \pm 0.10$ fm. This result obtained with no free parameters is 
in very good agreement with
the experimental value of 
$^4a=6.35\pm 0.02$ fm \cite{dilg}.

\vspace{1cm}
\noindent
{\large\bf Acknowledgements}

\noindent
We thank David Kaplan for extensive discussions.
Discussions with 
Ben Bakker, Joe Carlson, Vitaly Efimov, Jim Friar, Walter Gl\"ockle, 
and Martin Savage are also acknowledged.
UvK is grateful to Justus Koch for hospitality at NIKHEF where part of this 
work was carried out.
This research was supported in part by the DOE
grants DOE-ER-40561 (PFB) and DE-FG03-97ER41014 (UvK).


\vspace{1cm}
                  
\begin{thebibliography}{50}
\bibitem{bira0} 
 U. van Kolck, Washington preprint DOE/ER/41014-22-N97 (hep-ph/9707228), 
               to appear in the ``Proceedings of the 6th Conference on the 
               Intersections of Particle and Nuclear Physics'', 
               R.E. Mischke and T.W. Donnelly (editors), AIP Press.
\bibitem{weinberg} 
 S. Weinberg, {\it Phys. Lett.} {\bf B251} (1990) 288;
              {\it  Nucl. Phys.} {\bf B363} (1991) 3.
\bibitem{bira1} 
 C. Ord\'{o}\~{n}ez and U. van Kolck, {\it Phys. Lett.} {\bf B291} (1992) 459; 
 U. van Kolck, {\it Phys. Rev.} {\bf C49} (1994) 2932.
\bibitem{bira2}
 C. Ord\'{o}\~{n}ez, L. Ray, and U. van Kolck, 
               {\it Phys. Rev. Lett.} {\bf 72} (1994) 1982; 
               {\it Phys. Rev.} {\bf C53} (1996) 2086.
\bibitem{david1} 
 D.B. Kaplan, M.J. Savage, and M.B. Wise, 
               {\it Nucl. Phys.} {\bf B478} (1996) 629.
\bibitem{david2}
 D.B. Kaplan, {\it Nucl. Phys.} {\bf B494} (1997) 471.
\bibitem{lots}
 T.D. Cohen, {\it Phys. Rev.} {\bf C55} (1997) 67;
 D.R. Phillips and T.D. Cohen, {\it Phys. Lett.} {\bf B390} (1997) 7;
 K.A. Scaldeferri, D.R. Phillips, C.-W. Kao, and T.D. Cohen, 
      Maryland preprint UMD-PP-97-053 (nucl-th/9610049);
 M. Luke and A. Manohar, {\it Phys. Rev.} {\bf D55} (1997) 4129;
 G.P. Lepage, nucl-th/9706029;
 D.R. Phillips, S.R. Beane, and T.D. Cohen, Maryland preprints 
                         UMD-PP-97-119 (hep-th/9706070)
                         and UMD-PP-98-024 (nucl-th/9709062).
\bibitem{weinbergqp}
 S. Weinberg, {\it Phys. Rev.} {\bf 130} (1963) 776. 
\bibitem{efimov}
 V. Efimov, {\it Phys. Rev.} {\bf C47} (1993) 1876.
\bibitem{amante}
 R.D. Amado, in ``Elementary Particle Physics and Scattering Theory,
 Brandeis 1967'', M. Chr\'etien and S.S. Schweber (editors), Vol.2,
 Gordon and Breach, 1970.
\bibitem{nijm}
 J.J. de Swart, C.P.F. Terheggen, and V.G.J. Stoks, nucl-th/9509032.
\bibitem{jerry}
 G.A. Miller, B.M.K. Nefkens, and I. \v{S}laus, {\it Phys. Rep.} {\bf 194}
 (1990) 1.
\bibitem{skorny}
 G.V. Skorniakov and K.A. Ter-Martirosian, {\it Sov. Phys. JETP} {\bf 4} 
 (1957) 648.
\bibitem{recipes}
 ``Numerical Recipes in C, The Art of Scientific Computing'',
 W.H. Press, S.A. Teukolsky, W.T. Vetterling, and B.P. Flannery,
 Cambridge U. Press, 1992.
\bibitem{dilg}
 W. Dilg, L. Koester, and W. Nistler, {\it Phys. Lett.} {\bf B36} (1971) 208.
\end{thebibliography}
\end{document}